\documentclass{article}
\usepackage{spconfa4,amsmath,graphicx}
\usepackage{booktabs}
\usepackage{multirow}
\usepackage{tabularx}
\usepackage{url}

\title{Mixture-Constrained Max Pooling Improves Separation-Based \\ Bird Species Classification}
\name{%
  \begin{tabular}[t]{c}
    Yuzhu Wang$^1$, Kalle Lahtinen$^1$, Patrik Lauha$^{2,3}$, Shiqi Zhang$^1$, \\
    Panu Somervuo$^2$, Otso Ovaskainen$^3$, Tuomas Virtanen$^1$
  \end{tabular}%
}
\address{$^{1}$Signal Processing Research Center, Tampere University, Tampere, Finland \\
$^{2}$ Faculty of Biological and Environmental Sciences, University of Helsinki, Helsinki, Finland \\
$^{3}$ Department of Biological and Environmental Science, University of Jyväskylä, Jyväskylä, Finland \\
}

\begin{document}
\ninept
\maketitle
\begin{abstract}
Bird species classification from field recordings remains challenging due to overlapping vocalizations and incomplete species labels. We study source separation as a preprocessing for bird species classification to improve multi-species detection. Specifically, we employ an ensemble of two separators, FTRNN and TF-Locoformer, both trained with mixture invariant training~(MixIT). To address the false positive gain caused by separation errors in separated outputs, we propose mixture-constrained max pooling~(MCM), which clips the predicted probability from each separated channel based on the corresponding species probability in the original mixture. The classifier is applied to each separated output and the original mixture independently, and MCM aggregates the predictions into a final per-species probability. Experiments on two real-world datasets show that the ensemble outperforms individual separators and MCM outperforms standard max pooling across multiple metrics, and reveal that separation leads to both true positive gain for present species and false positive gain for absent species.
\end{abstract}
\section{Introduction}
\label{sec:intro}

Automatic bird species classification is central to passive 
acoustic monitoring for ecological surveys~\cite{darras2018meta-analysis,hudson2017role, xie2024automatic}, but species recognition from field recordings remains challenging. Such recordings typically contain overlapping vocalizations from multiple species and diverse environmental noises~\cite{adavanne2017convolutional, miyaguchi2025distilling}. Training data are also largely weakly labeled, with annotations covering only the dominant species in each clip, while background species are often incompletely labeled~\cite{kahl2024overview}. 

Source separation has been explored to improve bird species classification from field recordings. For single-species separation of Golden-Cheeked Warbler, separators trained on site-specific data combining target-species recordings with real local background noise outperform training on generic birdsong data~\cite{sasek2024semiautomated}. 
However, this improvement in separation quality did not consistently translate into gains in downstream classification performance~\cite{sasek2024semiautomated}. 
A GAN-based spectrogram translation approach has been shown to effectively identify, denoise, and separate target sound sources from real-world recordings~\cite{wang2025animal}, but its training requires manually annotated pixel-wise paired spectrograms. Scaling supervised approaches to real-world soundscapes remains challenging, as obtaining clean single-species recordings for a large number of species is difficult in practice. Mixture invariant training~(MixIT)~\cite{wisdom2020unsupervised, wisdom2021sparse} opens another avenue for bioacoustic applications, which trains separation models on mixtures of mixtures without clean reference signals. A TDCN++ separator trained with MixIT, combined with an ensemble of EfficientNet-B0 classifiers, demonstrated consistent classification improvement across multiple real-world soundscape datasets~\cite{denton2022improving}. A key finding was that including the original mixture alongside separated channels for classification outperformed using separated channels alone.
\begin{figure}
\centering
\includegraphics[width=0.99\linewidth]{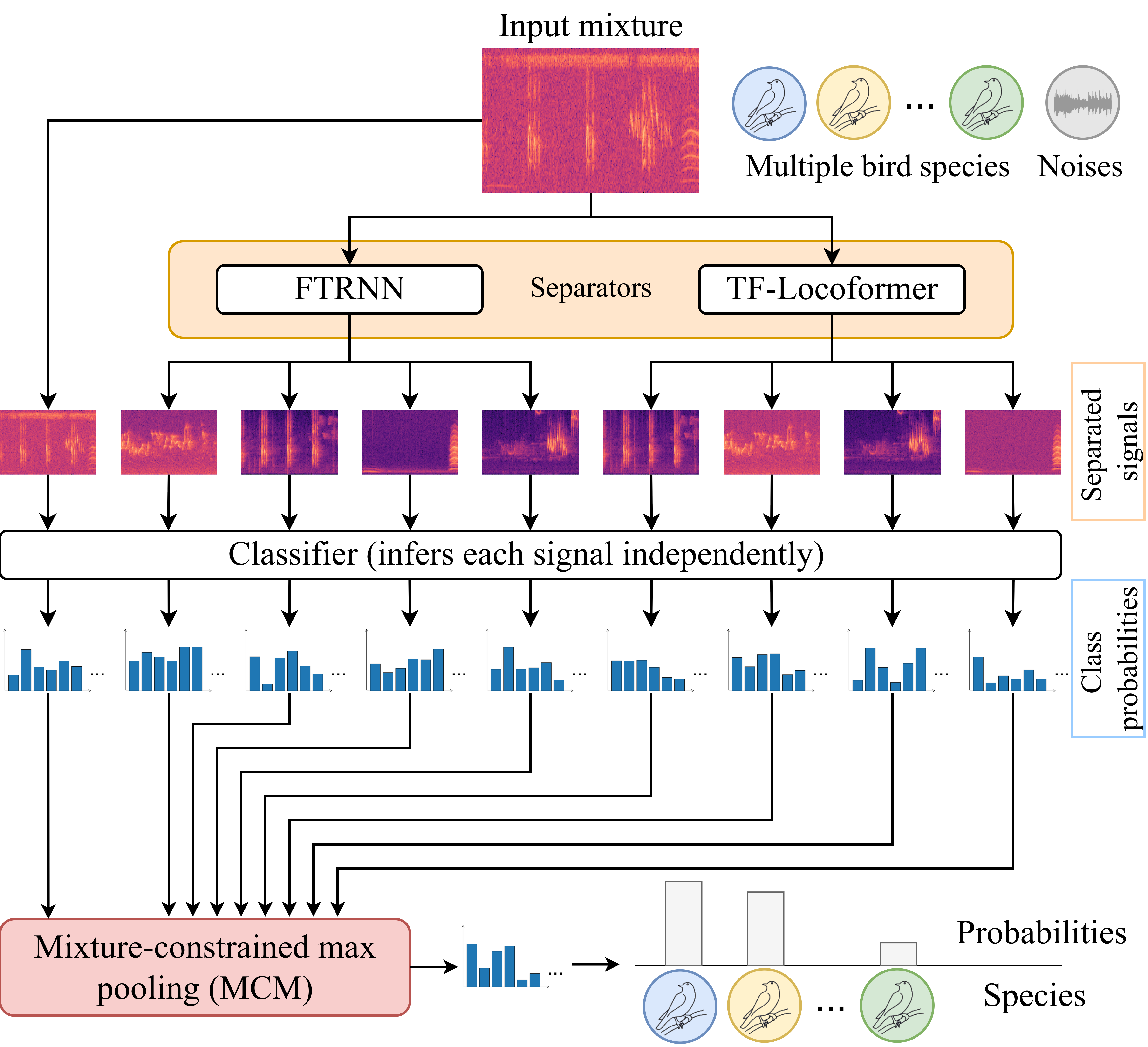}
\vspace{-2ex}
\caption{Visualization of the proposed system.}
\label{fig:system}
\vspace{-3ex}
\end{figure}

Separation can isolate species masked in the original mixture into individual output channels to boost their predicted probabilities (true positive gain). At the same time, separation may introduce source leakage and artifacts, inflating predicted probabilities for species absent from the recording (false positive gain). Aggregating predictions via max pooling over separated channels and the original mixture as used in~\cite{denton2022improving} effectively captures the true positive gain from separation, while the false positive gain problem remains overlooked. This bidirectional effect (true/false positive gain) of separation on classification performance has not been systematically studied, and designing aggregation strategies that account for both effects remains an open problem.

In this paper, we propose a separation-classification system that follows the same MixIT-based training paradigm as~\cite{denton2022improving}. 
As a first contribution, we propose mixture-constrained max pooling~(MCM), which limits the contribution of each separated channel based on the corresponding species probability in the original mixture, capturing true positive gain while suppressing false positive gain. 
As a second contribution, we introduce an ensemble of two separators. In the separate-then-classify paradigm, the classifier independently infers each output channel and the aggregation score is invariant to the ordering of output channels across separators, eliminating the need for permutation alignment and making multi-separator ensembling naturally feasible. We employ FTRNN~\cite{wang2025multi} and TF-Locoformer~\cite{saijo2024tf} as the two separators, whose architectural differences lead to complementary classification performance. 
Our experimental findings are threefold. (1) MCM outperforms max pooling across multiple metrics on both evaluation datasets. (2) A detailed analysis of the true and false positive gain of separation on classification provides insight into the bidirectional impact of separation on classification performance. (3) Separators trained on the same data show substantial performance differences, and ensembling consistently outperforms single-separator systems across multiple classification metrics, demonstrating complementary gains.

\section{Methods}
\label{sec:methods}
Our system detects bird species present in an input audio (3-second clips in our experiments), comprising two source separation models, a pre-trained bird species classifier, and an aggregation strategy using separation outputs, as illustrated in Fig.~\ref{fig:system}.
\vspace{-1ex}
\subsection{Separator}
\label{ssec:separation}

We employ two well-established separation models from the speech separation literature. FTRNN~\cite{wang2025multi} is an LSTM-based model operating in the time-frequency domain with full-band and subband modules. 
TF-Locoformer~\cite{saijo2024tf} is a transformer-based dual-path model in which self-attention captures global temporal and spectral dependencies while convolutional feed-forward networks handle local context, demonstrating competitive performance across multiple speech separation benchmarks.
Both separation models are trained using MixIT~\cite{wisdom2020unsupervised}.
Both models are configured to produce four output channels and trained with scale-invariant signal-to-distortion 
ratio (SI-SDR) loss~\cite{LeRoux2018a}.
FTRNN~\cite{wang2025multi} is configured with an embedding dimension of $128$, $6$ repeated blocks, and $256$ hidden units per direction in bi-directional long short-term memory (BLSTM) layers, scaled up from the original two-source to four-source configuration. TF-Locoformer uses the medium~(M) configuration~\cite{saijo2024tf}.
\vspace{-1ex}
\subsection{Classifier}
\label{ssec:classifier}

For bird species classification we use convolutional neural networks trained with Bird Sounds Global pipeline~\cite{lauha2025bsg} to classify Finnish and Madagascan bird species from passive acoustic monitoring data. Both models analyze audio in 3-second clips and output confidence scores for each species to be present in the clip. 
The classification models are available at Github: \url{https://github.com/plauha/BSG_classifier_builder}. For Finnish birds we use model v4.4, and for Madagascan birds model v3.

\begin{figure}
\centering
\includegraphics[width=\linewidth]{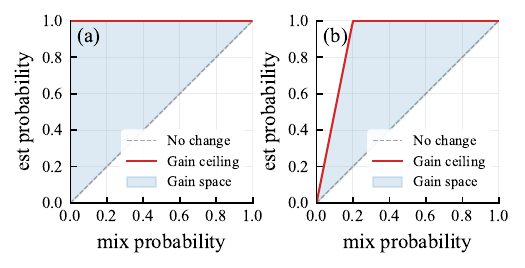}
\vspace{-5ex}
\caption{Predicted probability of a species from the mixture ($x$-axis) and from a separated output ($y$-axis). The gain ceiling defines the maximum predicted probability contributed by a separated output, and the shaded gain space indicates the region where the final predicted probability may fall. (a) Standard max pooling. (b) Proposed mixture-constrained max pooling with $\tau = 4.0$.}
\label{fig:mcm}
\vspace{-2ex}
\end{figure}
\vspace{-1ex}
\subsection{Classification with Separated Outputs}
\label{ssec:inference}
Given the classifier predictions from the original mixture and all separated outputs, the final predicted probability for each species is obtained by aggregating these predictions.
Max pooling over the original mixture and all separated outputs has been shown to improve classification performance~\cite{denton2022improving}. However, separated outputs may contain source leakage, separation artifacts, and residual noise, which may inflate predicted probabilities for certain species. When a species has a very small probability in the mixture yet a high probability in a separated output, such a large gain is more likely attributable to these interference components than to a real vocalization. To reduce the risk of such false positive gain, we propose mixture-constrained max pooling (MCM), which limits the contribution of each separated output based on the corresponding species probability in mixture. For each species $s$, the predicted probability from a separated output $p_{\text{est},s}$ is clipped according to
\begin{equation}
p'_{\text{est},s} = \begin{cases} (1 + \tau) \cdot p_{\text{mix},s} & 
\text{if } \frac{p_{\text{est},s} - p_{\text{mix},s}}{p_{\text{mix},s} + \epsilon} > \tau \\ 
p_{\text{est},s} & \text{otherwise} \end{cases},
\label{eq:mcm}
\end{equation}
where $p_{\text{mix},s}$ is the predicted probability of species 
$s$ from the mixture, $\tau \geq 0$ is a threshold, and $\epsilon$ is a small constant for numerical stability. The final predicted probability is taken as the maximum of $p_{\text{mix},s}$ and all clipped $p'_{\text{est},s}$ across separated outputs. 

As illustrated in Fig.~\ref{fig:mcm}, with standard max pooling, the maximum predicted probability from separated outputs is always $1.0$. With MCM, when $\tau = 0$, the probability increase from separation is fully suppressed, as any $p_{\text{est},s} > p_{\text{mix},s}$ is clipped back to $p_{\text{mix},s}$. As $\tau$ increases, the slope of the gain ceiling increases. 
\vspace{-1ex}
\section{Experimental settings}
\label{sec-sxperimental-settings}
\vspace{-1ex}
\subsection{Separation Datasets}
\label{ssec-sep-datasets}
The separation training data consist of bird vocalizations sourced from the Xeno-Canto archive~\cite{xenocanto} by downloading recordings originating from locations defined by a geographical coordinate bounding box (latitude range $[50.0, 72.0]$, longitude range $[-3.0, 39.0]$). Each recording is segmented into 3-second clips with 1-second overlap between consecutive windows and analyzed with the pre-trained bird species classifier in \ref{ssec:classifier}. Predictions with $>0.05$ classification for each 3-second clip within the recording are stored for signal-to-noise ratio (SNR) estimation. SNR is computed for each recording as the relation of energies in active and inactive clips within each recording. A clip is defined as active if the maximum classification confidence (the classification model's certainty of a bird species vocalization presence in the analyzed clip) for any bird species is $>0.5$. At maximum, the first 1,800 seconds of the recordings were used for computing the SNR estimation. Clips are retained only if the estimated SNR of a recording exceeds $30$~dB and the confidence score of the pre-trained bird classifier is $\geq0.8$. After filtering, the extracted corpus comprises 153 species, 34,681 clips, and approximately 28.9 hours of audio, with at least 10 clips per species. The corpus is randomly partitioned into training ($80\%$) and validation ($20\%$) pools at the recording level, ensuring that clips from the same original recording appear in only one split.

Training data are generated on-the-fly following the mixture-of-mixtures (MoM) protocol~\cite{denton2022improving}. At each iteration, two reference mixtures are sampled from the training pool, each formed by randomly mixing one or two clips, with no species shared within or between the two reference mixtures. The two reference mixtures are summed to form the MoM input. To extend the original MoM protocol and improve the noise suppression performance of the separation models, additional noise is added to MoM. With probability $0.8$, real-world environmental noise selected from field recordings collected with autonomous recorders and manually annotated to contain no bird sounds~\cite{lauha2025bsg} is added at an SNR uniformly sampled from $[0, 10]$~dB. With probability $0.5$, white Gaussian noise is added at an SNR uniformly sampled from $[20, 30]$~dB. When multiple bird vocalization clips are present, SNR is computed based on their mean power. To enable the separation models to handle single-source mixtures, with probability $0.1$, one randomly selected reference mixture is replaced by a zero signal. All generated signals are single-channel with a sampling frequency of $16$~kHz.
\begin{table}[t]
  \centering
  \caption{SI-SDRi (dB) of separation models evaluated on mixtures 
  with varying numbers of sources.}
  \label{tab:separation_results}
  \setlength{\tabcolsep}{4pt}
  \begin{tabularx}{\columnwidth}{Xccccccc}
    \toprule
    \multirow{2}{*}[-0.5ex]{Model} & \multirow{2}{*}[-0.5ex]{\shortstack{\#Param\\(M)}} & 
    \multicolumn{4}{c}{Sources in mixture} & \multirow{2}{*}[-0.5ex]{Avg} \\
    \cmidrule(lr){3-6}
     & & 1 & 2 & 3 & 4 & \\
    \midrule
    FTRNN             & 10.3 & 12.1 & 12.3 & 8.0 & 4.8 & 9.3 \\
    TF-Locoformer (M) & 15.0 & 11.9 & 9.0  & 4.7 & 1.8 & 6.9 \\
    \bottomrule
  \end{tabularx}
  \vspace{-2ex}
\end{table}
\vspace{-1ex}
\subsection{Evaluation Datasets}
\label{ssec-eval-datasets}
We evaluated the effect of sound separation on classification performance with clips from passive acoustic monitoring datasets collected with AudioMoth recording devices through international Lifeplan project~\cite{somervuo2025soundscape, hardwick2024lifeplan} and labeled on the BSG portal~\cite{lauha2025bsg}. 

The dataset of Finnish birds includes 4,600 3-second clips that contain 53 bird species (44 of which occur in more than five recordings), on average 1.73 species per clip, with 1,935 clips (42.1\%) containing at most one species and 2,665 clips (57.9\%) containing more than one. The dataset was collected from a set of more than 200,000 1-minute field recordings. The 1-minute recordings were analyzed with the Finnish classification model and a subset of recordings that were most likely to contain multiple vocalizations of more than one bird species were selected for manual review. An experienced bird expert reviewed the recordings and produced the ground truth labels for them. Finally, the 3-second clips were extracted by splitting each 1-minute recording into 20 consecutive clips. 

The dataset of Madagascan birds includes 2,215 3-second clips that contain 87 bird species (58 of which occur in more than five recordings), on average 1.46 species per clip, with 1,193 clips (53.9\%) containing at most one species and 1,022 clips (46.1\%) containing more than one. The original field recordings were analyzed with BirdNET classifier~\cite{kahl2021birdnet} and 20-second clips with likely bird vocalizations (but not specifically multiple overlapping species as in the Finnish dataset) were selected for manual annotation by an experienced bird expert. 3-second clips were extracted around the parts that were annotated to contain bird vocalizations and this data set was randomly split into training and validation sets used for the Madagascan classification model~\cite{lauha2025bsg}. 
\vspace{-1ex}
\subsection{Evaluation Metrics}
\label{ssec-metrics}
We evaluate separation quality using the SI-SDR improvement (SI-SDRi)~\cite{LeRoux2018a}. Classification performance is assessed using 
class-averaged mean average precision (CMAP)~\cite{kahl2019overview}, label-weighted label-ranking 
average precision (lwlrap)~\cite{fonseca2019audio}, and area under the ROC curve (AUC). CMAP and AUC 
are computed over species with more than five observations in a given dataset, 
as species with few observations result in unreliable metric values.
\begin{table*}[t]
  \centering
  \caption{Classification performance on the Finland and Madagascar 
  evaluation datasets.}
  \vspace{0.5ex}
  \label{tab:classification_results}
  \setlength{\tabcolsep}{7pt}
  \begin{tabularx}{\textwidth}{clXlcccccc}
    \toprule
    \multirow{2}{*}{\#} & \multirow{2}{*}{System} & 
    \multirow{2}{*}{Separator} & 
    \multirow{2}{*}{\shortstack{Aggregation\\strategy}} &
    \multicolumn{3}{c}{Finland} & 
    \multicolumn{3}{c}{Madagascar} \\
    \cmidrule(lr){5-7} \cmidrule(lr){8-10}
    & & & & CMAP & lwlrap & AUC & CMAP & lwlrap & AUC \\
    \midrule
    0 & Baseline  & --  & --                              & 0.448 & 0.743 & 0.841 & 0.531 & \textbf{0.724} & 0.894 \\
    1 & Ensemble  & FTRNN + TF-Locoformer (M) & MCM ($\tau$=50)  & \textbf{0.485} & \textbf{0.756} & \textbf{0.883} & \textbf{0.553} & 0.723 & \textbf{0.908} \\
    \midrule
    2 & Single    & FTRNN                     & MCM ($\tau$=50)  & 0.473 & 0.740 & 0.879 & 0.549 & 0.717 & 0.904 \\
    3 & Single    & TF-Locoformer (M)         & MCM ($\tau$=50)  & 0.477 & \textbf{0.762} & 0.862 & 0.543 & 0.727 & 0.896 \\
    4 & Ensemble  & FTRNN + TF-Locoformer (M) & Max pooling      & 0.477 & 0.744 & 0.882 & 0.551 & 0.714 & 0.905 \\
    5 & Ensemble  & FTRNN + TF-Locoformer (M) & MCM ($\tau$=100) & 0.484 & 0.751 & \textbf{0.886} & 0.552 & 0.720 & 0.906 \\
    6 & Ensemble  & FTRNN + TF-Locoformer (M) & MCM ($\tau$=75)  & 0.484 & 0.753 & 0.885 & 0.552 & 0.721 & 0.907 \\
    7 & Ensemble  & FTRNN + TF-Locoformer (M) & MCM ($\tau$=25)  & \textbf{0.486} & \textbf{0.762} & 0.880 & \textbf{0.554} & \textbf{0.728} & \textbf{0.909} \\
    \bottomrule
  \end{tabularx}
  \vspace{-2ex}
\end{table*}
\vspace{-1ex}
\section{Results and Discussions}
\label{sec-Experimental-Results}
\vspace{-1ex}
\subsection{Separation Results}
\label{ssec-sep-results}
The separation models are evaluated on four test subsets, each containing 2,000 samples and corresponding to mixtures with one to four species. Each sample is generated by randomly sampling clips from distinct species in the validation pool and summing them to form a clean mixture, following the same noise addition procedure as training with real environmental noise at an SNR uniformly sampled from $[0, 10]$~dB and white Gaussian noise at an SNR uniformly sampled from $[20, 30]$~dB, both applied with probability $1.0$. 
Each separator produces $4$ output channels regardless of the number of 
sources $K$ in the input mixture. SI-SDR is computed over all $\binom{4}{K}$ subsets of $K$ output channels. For each subset, the optimal permutation between the $K$ selected channels and the $K$ reference signals is determined by maximizing the mean SI-SDR. The maximum mean SI-SDR across all subsets is used as the final score.

Table~\ref{tab:separation_results} reports SI-SDRi of the two separation models. FTRNN outperforms TF-Locoformer~(M) across all conditions despite having fewer parameters ($10.3$~M vs.\ $15.0$~M). Both models show decreasing SI-SDRi as the number of sources increases, with performance dropping more steeply for TF-Locoformer~(M), from $11.9$~dB with one source to $1.8$~dB with four sources, compared to $12.1$~dB to $4.8$~dB for FTRNN.
\begin{table*}[t]
  \centering
  \caption{True positive gain (TPG) and false positive gain (FPG) on the Finland and Madagascar evaluation datasets. Statistics are computed over all species-clip pairs. "Clipped by MCM" reports the number and fraction of clipped predictions attributable to each class.}
  \vspace{1ex}
  \label{tab:tp_fp_analysis}
  \setlength{\tabcolsep}{5pt}
  \renewcommand{\arraystretch}{1.15}
  \begin{tabularx}{\textwidth}{Xlrr ccc ccc ccc}
    \toprule
    \multirow{2}{*}[-0.6ex]{Testset}
      & \multirow{2}{*}[-0.6ex]{Class}
      & \multirow{2}{*}[-0.6ex]{$N$}
      & \multirow{2}{*}[-0.6ex]{\%}
      & \multicolumn{3}{c}{$p_{\text{mix}}$}
      & \multicolumn{3}{c}{$\max(p_{\text{est}}) - p_{\text{mix}}$}
      & \multicolumn{3}{c}{Clipped by MCM (Num / \%)} \\
    \cmidrule(lr){5-7}\cmidrule(lr){8-10}\cmidrule(lr){11-13}
      & & & & Mean & Med. & STD
              & Mean & Med. & STD
              & $\tau{=}25$ & $\tau{=}50$ & $\tau{=}75$ \\
    \midrule
    \multirow{2}{*}{Finland}
      & TPG & 6,659  &  2.7 & 0.262 & 0.094 & 0.315 & 0.182 & 0.100 & 0.204 & 375/1.6 & 183/1.5 & 109/1.3 \\
      & FPG & 196,369 & 80.5 & 0.004 & 0.001 & 0.017 & 0.013 & 0.003 & 0.045 & 22,508/98.4 & 12,074/98.5 & 8,339/98.7 \\
    \addlinespace
    \multirow{2}{*}{Madagascar}
      & TPG &   2,493 &  1.3 & 0.417 & 0.310 & 0.369 & 0.163 & 0.088 & 0.187 & 54/0.2 & 23/0.2 & 12/0.1 \\
      & FPG & 170,378 & 88.4 & 0.010 & 0.002 & 0.035 & 0.027 & 0.008 & 0.063 & 24,418/99.8 & 12,633/99.8 & 8,385/99.9 \\
    \bottomrule
  \end{tabularx}
  \vspace{-2ex}
\end{table*}
\vspace{-1ex}
\subsection{Classification Results and Ablation Experiments}
\label{ssec-clf-results}
Table~\ref{tab:classification_results} reports classification performance on the Finland and Madagascar test sets. 
The baseline applies the classifier directly to the mixture. For separation-based systems, the classifier is applied to all separated channels and the original mixture, yielding 9 prediction vectors for the ensemble (4 + 4 + 1) and 5 for single-separator (4 + 1) configurations, aggregated via MCM or max pooling.
The value of $\tau$ used in the experiments was determined through pre-experiments on the validation dataset.

The proposed system (Row~\#~1) improves over the baseline (Row~\#~0) on all metrics on the Finland dataset, with CMAP increasing from $0.448$ to 
$0.485$, lwlrap from $0.743$ to $0.756$, and AUC from $0.841$ to $0.883$. On the Madagascar dataset, CMAP increases from $0.531$ to $0.553$, while AUC shows only marginal improvement and lwlrap decreases slightly.
Comparing Row~\#~1 with Rows~\#~2--3, the proposed system outperforms both individual separators on most metrics. The two single-separator systems show notable performance differences despite using the same training data, with TF-Locoformer~(M) achieving higher lwlrap than FTRNN on both datasets ($0.762$ vs.\ $0.740$ on Finland, $0.727$ vs.\ $0.717$ on Madagascar), while FTRNN achieves higher AUC on both datasets ($0.879$ vs.\ $0.862$ on Finland, $0.904$ vs.\ $0.896$ on Madagascar). This suggests that different separator architectures produce different inductive biases, and combining them yields complementary gains.
Comparing the two aggregation strategies, MCM (Row~\#~1) outperforms max pooling (Row~\#~4) on all metrics across both datasets.
Rows~\#~5--7 show the effect of varying $\tau$. Classification metrics do not change monotonically with $\tau$.
Comparing Table~\ref{tab:classification_results} and 
Table~\ref{tab:separation_results}, separation quality measured by 
SI-SDRi does not directly correlate with classification performance, as 
higher SI-SDRi does not consistently lead to improvement across all 
classification metrics.
\vspace{-1ex}
\subsection{True/False Positive Gain Analysis}
\label{ssec-Falseimprovement}
To examine the effect of separation and MCM on classification, we analyzed all species-clip pairs in the evaluation datasets using the 
proposed system (Row~\#~1 in Table~\ref{tab:classification_results}). The Finland test set comprises 53 species and 4,600 clips (243,800 species-clip pairs), and the Madagascar test set comprises 87 species and 2,215 clips (192,705 species-clip pairs). For each species-clip pair, $\max (p_{\mathrm{est},s})$ and $p_{\mathrm{mix},s}$ denote the maximum predicted probability for species $s$ across all separated channels and the predicted probability on the original mixture, respectively. A species-clip pair is labeled true positive gain~(TPG) if the species is present and $\max (p_{\mathrm{est},s}) > p_{\mathrm{mix},s}$, and false positive gain~(FPG) if the species is absent and $\max (p_{\mathrm{est},s}) > p_{\mathrm{mix},s}$. 
True and false negative gains are excluded from this analysis. When max pooling is applied and $\max (p_{\mathrm{est},s}) \leq p_{\mathrm{mix},s}$, the final predicted probability remains $p_{\mathrm{mix},s}$ regardless of $p_{\mathrm{est},s}$. This implies that max pooling inherently eliminates the effect of separation on true and false negative gains.

As shown in Table~\ref{tab:tp_fp_analysis}, the proportion of TPG pairs is small on both test sets ($2.7\%$ on Finland and $1.3\%$ on Madagascar), while FPG pairs account for the majority ($80.5\%$ and $88.4\%$), with the remaining $16.8\%$ and $10.3\%$ corresponding to pairs where separation has no effect on the final predicted probability due to max pooling. 
Although FPG pairs are numerous, most exhibit small values of $\max(p_{\mathrm{est},s}) - p_{\mathrm{mix},s}$, suggesting that their impact on classification is limited. These results indicate that separation benefits classification by substantially improving the predicted probability for a small number of present species. It results in a large number of FPG pairs at the same time, most of which exhibit only marginal probability changes. FPG pairs with large probability increases are more likely to decrease classification performance, and reducing their contribution is important for improving the separation-classification pipeline. 
This is consistent with the motivation of the proposed MCM. As illustrated in Fig.~\ref{fig:mcm}, MCM applies stronger suppression to predictions with smaller $p_{\mathrm{mix},s}$ and larger probability increases, targeting the region where harmful FPG pairs are most concentrated.
In Table~\ref{tab:tp_fp_analysis}, at $\tau{=}50$, $98.5\%$ and $99.8\%$ of all clipped predictions are attributed to FPG pairs on Finland and Madagascar, with only $1.5\%$ and $0.2\%$ affecting TPG pairs. Across all tested values of $\tau$, over $98\%$ of clipped predictions are attributed to FPG pairs, demonstrating that MCM consistently suppresses false positive gains while leaving true positive gains largely unaffected, and suggesting robustness to the choice of $\tau$.
\vspace{-2ex}
\section{Conclusions}
\label{sec-conclusions}
We studied separation-based bird species classification, proposing MCM as an aggregation strategy to suppress false positive gain and introducing a separator ensemble to improve classification performance. We found that separators with different architectures yield complementary classification performance, while higher separation quality does not necessarily lead to better classification. MCM consistently outperforms standard max pooling by using the mixture prediction as a prior to suppress harmful probability inflations in separated outputs. The true and false positive gain analysis shows that separation substantially boosts predicted probabilities for a small number of masked species, while simultaneously inflating predicted probabilities for a large number of absent species. Most of these cases involve only marginal probability increases, but a small fraction with large increases can harm classification performance. 
Our analysis reveals considerable room for improvement in the separate-then-classify pipeline, and future work could explore reducing the domain mismatch between separation and classification systems.

\newpage
\bibliographystyle{IEEEbib}
\bibliography{Bib_YWang}

\end{document}